\documentclass[aps,prc,preprint,superscriptaddress, showpacs]{revtex4-1}

\usepackage[dvips]{graphics,graphicx}
\usepackage{epsfig}
\usepackage{ctable}
\begin{document}

\title{High-spin ${\mu}$s isomeric states in $^{96}$Ag}

\author{A. D. Becerril}
\email {ana\_deliab@hotmail.com}

\affiliation{National Superconducting Cyclotron Laboratory, Michigan State University, East Lansing, MI 48824, USA}
\affiliation{Department of Physics and Astronomy, Michigan State University, East Lansing, MI 48824, USA}
\affiliation{Joint Institute for Nuclear Astrophysics, Michigan State University, East Lansing, MI 48824, USA}

\author{G.~Lorusso}
\affiliation{National Superconducting Cyclotron Laboratory, Michigan State University, East Lansing, MI 48824, USA}
\affiliation{Department of Physics and Astronomy, Michigan State University, East Lansing, MI 48824, USA}
\affiliation{Joint Institute for Nuclear Astrophysics, Michigan State University, East Lansing, MI 48824, USA}

\author{A. M.~Amthor}
\affiliation{National Superconducting Cyclotron Laboratory, Michigan State University, East Lansing, MI 48824, USA}
\affiliation{Department of Physics and Astronomy, Michigan State University, East Lansing, MI 48824, USA}
\affiliation{Joint Institute for Nuclear Astrophysics, Michigan State University, East Lansing, MI 48824, USA}

\author{T.~Baumann}
\affiliation{National Superconducting Cyclotron Laboratory, Michigan State University, East Lansing, MI 48824, USA}

\author{D.~Bazin}
\affiliation{National Superconducting Cyclotron Laboratory, Michigan State University, East Lansing, MI 48824, USA}

\author{J.~S.~Berryman}
\affiliation{National Superconducting Cyclotron Laboratory, Michigan State University, East Lansing, MI 48824, USA}
\affiliation{Department of Chemistry, Michigan State University, East Lansing, MI 48824, USA}

\author{B.~A.~Brown}
\affiliation{National Superconducting Cyclotron Laboratory, Michigan State University, East Lansing, MI 48824, USA}
\affiliation{Department of Physics and Astronomy, Michigan State University, East Lansing, MI 48824, USA}
\affiliation{Joint Institute for Nuclear Astrophysics, Michigan State University, East Lansing, MI 48824, USA}

\author{H. L.~Crawford}
\affiliation{National Superconducting Cyclotron Laboratory, Michigan State University, East Lansing, MI 48824, USA}
\affiliation{Department of Chemistry, Michigan State University, East Lansing, MI 48824, USA}

\author{A.~Estrade}
\affiliation{National Superconducting Cyclotron Laboratory, Michigan State University, East Lansing, MI 48824, USA}
\affiliation{Department of Physics and Astronomy, Michigan State University, East Lansing, MI 48824, USA}
\affiliation{Joint Institute for Nuclear Astrophysics, Michigan State University, East Lansing, MI 48824, USA}

\author{A.~Gade}
\affiliation{National Superconducting Cyclotron Laboratory, Michigan State University, East Lansing, MI 48824, USA}
\affiliation{Department of Physics and Astronomy, Michigan State University, East Lansing, MI 48824, USA}

\author{T.~Ginter}
\affiliation{National Superconducting Cyclotron Laboratory, Michigan State University, East Lansing, MI 48824, USA}

\author{C. J.~Guess}
\affiliation{National Superconducting Cyclotron Laboratory, Michigan State University, East Lansing, MI 48824, USA}
\affiliation{Department of Physics and Astronomy, Michigan State University, East Lansing, MI 48824, USA}  
\affiliation{Joint Institute for Nuclear Astrophysics, Michigan State University, East Lansing, MI 48824, USA}

\author{M.~Hausmann}
\affiliation{National Superconducting Cyclotron Laboratory, Michigan State University, East Lansing, MI 48824, USA}

\author{G.~W.~Hitt}
\altaffiliation[Current address: ]{Khalifa University of Science, Technology, and Research, Abu Dhabi Campus, P.O. Box 127788, Abu Dhabi, UAE}
\affiliation{National Superconducting Cyclotron Laboratory, Michigan State University, East Lansing, MI 48824, USA}
\affiliation{Department of Physics and Astronomy, Michigan State University, East Lansing, MI 48824, USA}
\affiliation{Joint Institute for Nuclear Astrophysics, Michigan State University, East Lansing, MI 48824, USA}

\author{P. F.~Mantica}
\affiliation{National Superconducting Cyclotron Laboratory, Michigan State University, East Lansing, MI 48824, USA}
\affiliation{Department of Chemistry, Michigan State University, East Lansing, MI 48824, USA}

\author{M.~Matos}
\affiliation{National Superconducting Cyclotron Laboratory, Michigan State University, East Lansing, MI 48824, USA}
\affiliation{Joint Institute for Nuclear Astrophysics, Michigan State University, East Lansing, MI 48824, USA}

\author{R.~Meharchand}
\affiliation{National Superconducting Cyclotron Laboratory, Michigan State University, East Lansing, MI 48824, USA}
\affiliation{Department of Physics and Astronomy, Michigan State University, East Lansing, MI 48824, USA}
\affiliation{Joint Institute for Nuclear Astrophysics, Michigan State University, East Lansing, MI 48824, USA}

\author{K.~Minamisono}
\affiliation{National Superconducting Cyclotron Laboratory, Michigan State University, East Lansing, MI 48824, USA}

\author{F.~Montes}
\affiliation{National Superconducting Cyclotron Laboratory, Michigan State University, East Lansing, MI 48824, USA}
\affiliation{Joint Institute for Nuclear Astrophysics, Michigan State University, East Lansing, MI 48824, USA}

\author{G.~Perdikakis}
\affiliation{National Superconducting Cyclotron Laboratory, Michigan State University, East Lansing, MI 48824, USA}
\affiliation{Joint Institute for Nuclear Astrophysics, Michigan State University, East Lansing, MI 48824, USA}

\author{J.~Pereira}
\affiliation{National Superconducting Cyclotron Laboratory, Michigan State University, East Lansing, MI 48824, USA}
\affiliation{Joint Institute for Nuclear Astrophysics, Michigan State University, East Lansing, MI 48824, USA}

\author{M.~Portillo}
\affiliation{National Superconducting Cyclotron Laboratory, Michigan State University, East Lansing, MI 48824, USA}

\author{H.~Schatz}
\affiliation{National Superconducting Cyclotron Laboratory, Michigan State University, East Lansing, MI 48824, USA}
\affiliation{Department of Physics and Astronomy, Michigan State University, East Lansing, MI 48824, USA}
\affiliation{Joint Institute for Nuclear Astrophysics, Michigan State University, East Lansing, MI 48824, USA}

\author{K.~Smith}
\affiliation{National Superconducting Cyclotron Laboratory, Michigan State University, East Lansing, MI 48824, USA}
\affiliation{Department of Physics and Astronomy, Michigan State University, East Lansing, MI 48824, USA}
\affiliation{Joint Institute for Nuclear Astrophysics, Michigan State University, East Lansing, MI 48824, USA}

\author{ J.~Stoker}
\affiliation{National Superconducting Cyclotron Laboratory, Michigan State University, East Lansing, MI 48824, USA}
\affiliation{Department of Chemistry, Michigan State University, East Lansing, MI 48824, USA}

\author{A.~Stolz}
\affiliation{National Superconducting Cyclotron Laboratory, Michigan State University, East Lansing, MI 48824, USA}

\author{R. G. T.~Zegers}
\affiliation{National Superconducting Cyclotron Laboratory, Michigan State University, East Lansing, MI 48824, USA}
\affiliation{Department of Physics and Astronomy, Michigan State University, East Lansing, MI 48824, USA}
\affiliation{Joint Institute for Nuclear Astrophysics, Michigan State University, East Lansing, MI 48824, USA}
\date{\today}

\begin{abstract}
The isomeric and ${\beta}$ decays of the $N = Z + 2$ nucleus $^{96}$Ag were investigated at NSCL. A cascade of ${\gamma}$-ray transitions originating from the de-excitation of a ${\mu}$s isomer was observed for the first time and was found in coincidence with two previously-known transitions with energies of 470 and 667 keV. The isomeric half-life was determined as 1.45(7) ${\mu}$s, more precise than previously reported. The existence of a second, longer-lived ${\mu}$s isomer, associated with a 743-keV ${\gamma}$ transition, is also proposed here. Shell model results within the $(p_{3/2}p_{1/2}f_{5/2}g_{9/2})$ model space, using the jj44b interaction, reproduced level energies and isomeric decay half-lives reasonably well.
\end{abstract}

\maketitle

The region of the chart of nuclides around $^{100}$Sn has been the focus of major experimental efforts in the last few decades. The low-energy structure of these nuclei is of great importance to confirm the closure of shells at $N = Z = 50$. New data on $^{100}$Sn and its closest neighbors allow testing of model spaces and effective interactions used in shell model calculations in this mass region. A particularly interesting phenomenon in $N \approx Z$ nuclei is the occurrence of high-spin isomers. These isomeric states appear due to an extra binding energy originating from the large attractive proton-neutron interaction in these configurations \cite{Kaneko2008, Hasegawa2005}.  A number of isomers have been predicted to exist in the region around $^{100}$Sn (e.g. \cite{Ogawa1983}), and several have been experimentally confirmed. Some remarkable examples of high-spin ${\gamma}$-decaying isomers are the 23/2$^{+}$ and 37/2$^{+}$ states in $^{95}$Ag \cite{Doring2003,Marginean2003}, 14$^{+}$ in $^{94}$Pd \cite{Marginean2003,Plettner2004}, and 12$^{+}$ in $^{98}$Cd \cite{Blazhev2004}. ${\beta}$- and ${\beta}$- delayed proton (${\beta}$p)-decaying isomers have also been found in the region, examples of which are the 21$^{+}$ level in $^{94}$Ag \cite{LaCommara2002, Mukha2004} and  the 25/2$^{+}$ level in $^{97}$Cd \cite{Lorusso2011}. In this work, we report on the observation of ${\gamma}$-decaying states in the $T_{z}=1$ nucleus $^{96}$Ag, with three proton holes and one neutron hole relative to $^{100}$Sn.

$^{96}$Ag is on the path of the astrophysical rapid proton capture (rp) process \cite{Schatz2006}. Isomers along the reaction path may affect proton capture rates significantly by creating a non-thermal population of levels. Thus, experimental information on isomers in $^{96}$Ag contributes to the understanding of the production of mass $A=96$ in the rp-process.

$^{96}$Ag was identified by Kurcewicz {\it et al.} \cite{Kurcewicz1982} using the $^{60}$Ni($^{40}$Ca,p3n)$^{96}$Ag reaction. A half-life of 5.1(4) s was deduced for this nucleus from the analysis of ${\beta}$-delayed protons, and spin and parity of 8$^{+}$ or 9$^{+}$ was suggested for the ground state. Later, Batist {\it et al.} \cite{Batist2003} identified two ${\beta}$-decaying states in $^{96}$Ag with half-lives of 4.40(6) and 6.9(6) s and tentative spins and parities of 8$^{+}$ and 2$^{+}$, respectively; however, none of them was assigned as the ground-state. Evidence for additional excited states in $^{96}$Ag was reported by Grzywacz {\it et al.} \cite{Grzywacz1997}, who discovered the existence of an isomeric state with a half-life of 0.7 (2) ${\mu}$s and observed two ${\gamma}$ transitions with energies of 470 keV and 667 keV originating from the  ${\gamma}$ decay of the isomer; however, statistics were not sufficient to construct a level scheme. Based on a shell model prediction in the proton-neutron (p$_{1/2}$, g$_{9/2}$) model space using an empirical interaction \cite{Gross1976}, Grzywacz {\it et al.} suggested the decay of a 15$^{+}$ or 13$^{-}$ isomer followed by a few ${\gamma}$ transitions. In this work, we report a new measurement of the lifetime of this isomer, and the observation of several new ${\gamma}$ transitions, one of which indicates the existence of a second, longer-lived ${\mu}$s isomeric state in $^{96}$Ag. 

A secondary ``cocktail'' beam containing $^{96}$Ag in its ground and isomeric states was produced at NSCL by fragmentation of a 120~MeV/nucleon $^{112}$Sn primary beam impinging a 195 mg/cm$^{2}$ thick $^{9}$Be target. Results on other nuclei that comprised the cocktail beam have been reported in Refs.~\cite{Lorusso2011} and \cite{Bazin2008}. The secondary beam containing $^{96}$Ag was selected with the A1900 Fragment Separator \cite{Morrissey2003}, utilizing a 40.6 mg/cm$^{2}$ achromatic Kapton degrader located at the intermediate focal plane. The A1900 momentum acceptance was limited to 1\%. The separation by energy loss and magnetic rigidity provided by the A1900 Fragment Separator is, however, not sufficient for decay studies with proton-rich beams. This is due to the low-momentum tails of much more intensely produced nuclei closer to stability, which overlap within the A1900 acceptance of the rare isotopes of interest. An additional filtering stage was therefore provided by the Radio Frequency Fragment Separator (RFFS) \cite{Bazin2009}. The RFFS reduced the intensity of unwanted species in the beam by about a factor of 200, resulting in an average total rate at the experimental station of 50 nuclei per second.

The experimental station was composed of the NSCL Beta Counting System (BCS) \cite{Prisciandaro2002} and sixteen high-purity Ge detectors from the Segmented Germanium Array (SeGA) \cite{Mueller2001}. The BCS consisted of three silicon PIN detectors used to perform particle identification through energy loss and time of flight measurements. Downstream of the PIN detectors was a 1-mm thick Double-sided Silicon Strip Detector (DSSD), with an effective segmentation of 1600 pixels. Ions were implanted into the DSSD and their decay products (positrons and protons) were identified and correlated within time and position gates. Downstream of the DSSD were six 1-mm thick Single-sided Silicon Strip Detectors (SSSDs) and a 1-cm thick planar germanium detector to veto light particles in the beam. 

The sixteen Ge detectors from SeGA were placed in close geometry around the DSSD to measure prompt and ${\beta}$-delayed ${\gamma}$ rays within a 12-${\mu}$s time window after a charged particle triggered the data acquisition. The absolute ${\gamma}$-detection efficiency of the array was 6.1\% at 1~MeV, as determined with calibrated radioactive sources of $^{56}$Co, $^{125}$Sb, $^{154}$Eu, and $^{155}$Eu.

\begin{figure}[htpb]
\centering
\includegraphics[width=8.0cm]{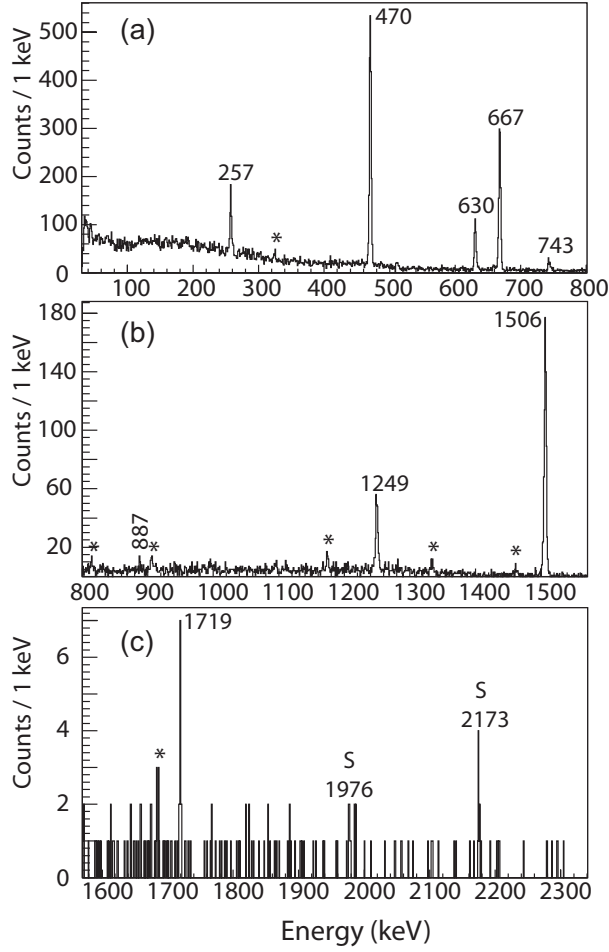}
\caption{ ${\gamma}$-ray spectrum collected between 1.2 and 12~${\mu}$s after an $^{96}$Ag implantation. Transitions belonging to the isomeric decay of $^{96}$Ag are labeled by their energies in keV. Lines marked with an asterisk are known background from neighboring $^{94}$Pd. Summing peaks are marked by the letter S.}
\label{fig:ggspectrum}
\end{figure}

The spectrum of ${\gamma}$ rays observed between 1.2 and 12~${\mu}$s after implantation of a $^{96}$Ag ion is shown in Fig. \ref{fig:ggspectrum}. The lower time limit of 1.2 ${\mu}$s was chosen to eliminate prompt X-rays and low-energy ${\gamma}$-rays produced by beam fragments interacting with the silicon detectors. Apart from the 470 keV and 667 keV transitions, which were previously reported by Grzywacz {\it et al.} \cite{Grzywacz1997}, all of the ${\gamma}$ transitions in Fig. \ref{fig:ggspectrum} were observed in the present experiment for the first time. 

\begin{figure}[htpb]
\includegraphics[width=12cm]{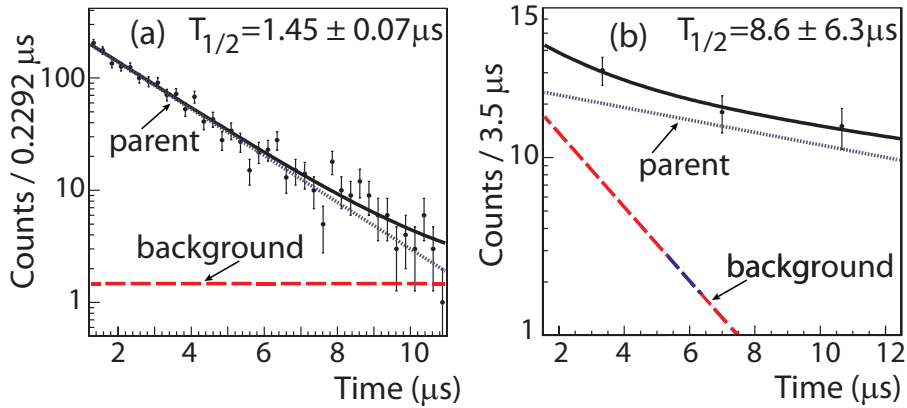}
\caption{(Color online) (a) Decay curve obtained by gating on the 257, 630, 667, and 1506-keV ${\gamma}$-ray transitions. The curve was fitted with a single exponential and constant background. (b) Decay curve obtained by gating on the 743-keV ${\gamma}$-ray transition, fitted with a single exponential decay and a time-dependent background.}
\label{fig:Ag96_isodecaycurve}
\end{figure}

The time distribution of each of the ${\gamma}$ lines of interest was analyzed to assign them to the decay of a specific isomer and to determine isomeric half-lives. The 257, 630, 667, and 1506-keV transitions decay with the same half-life. An OR of these energies was used to obtain the isomeric decay curve shown in Fig. \ref{fig:Ag96_isodecaycurve}(a). A fit to the decay curve taking into account a single exponential decay plus a constant background resulted in a half-life of 1.45(7) ${\mu}$s, more than two times longer than the previously-reported value of 0.7(2) ${\mu}$s \cite{Grzywacz1997}. The 743-keV transition, however, appears to decay with a longer half-life (See Fig. \ref{fig:Ag96_isodecaycurve}(b)). The fit to this decay curve needed to consider the time dependence of the background attributed to the contribution from the decay of the shorter-lived isomer. The low intensity of the 743-keV transition and the short 12 ${\mu}$s collection time results in large error bars in the half-life determination: 8.6(63)$~{\mu}$s. Nontheless, we interpret this as indication that the 743-keV transition corresponds to the decay of an additional, longer-lived microsecond isomer in $^{96}$Ag. 

The isomeric fractions were calculated following the procedure outlined by Daugas {\it et al.} \cite{Daugas2001}, considering the number of counts in the 630-keV and 667-keV peaks, the total number of implanted $^{96}$Ag ions, and the 475 ns flight time of the fragments from the target to the DSSD. An isomeric production fraction of 10.8(12)\% was estimated for the short-lived isomer in the fragmentation process. Similarly, by comparing the number of counts in the 743-keV peak to the total number of $^{96}$Ag implants, a 0.59(9)\% production of the long-lived isomer was estimated.

\begin{figure*}[htpb]
\centering
\includegraphics[width=16.0cm]{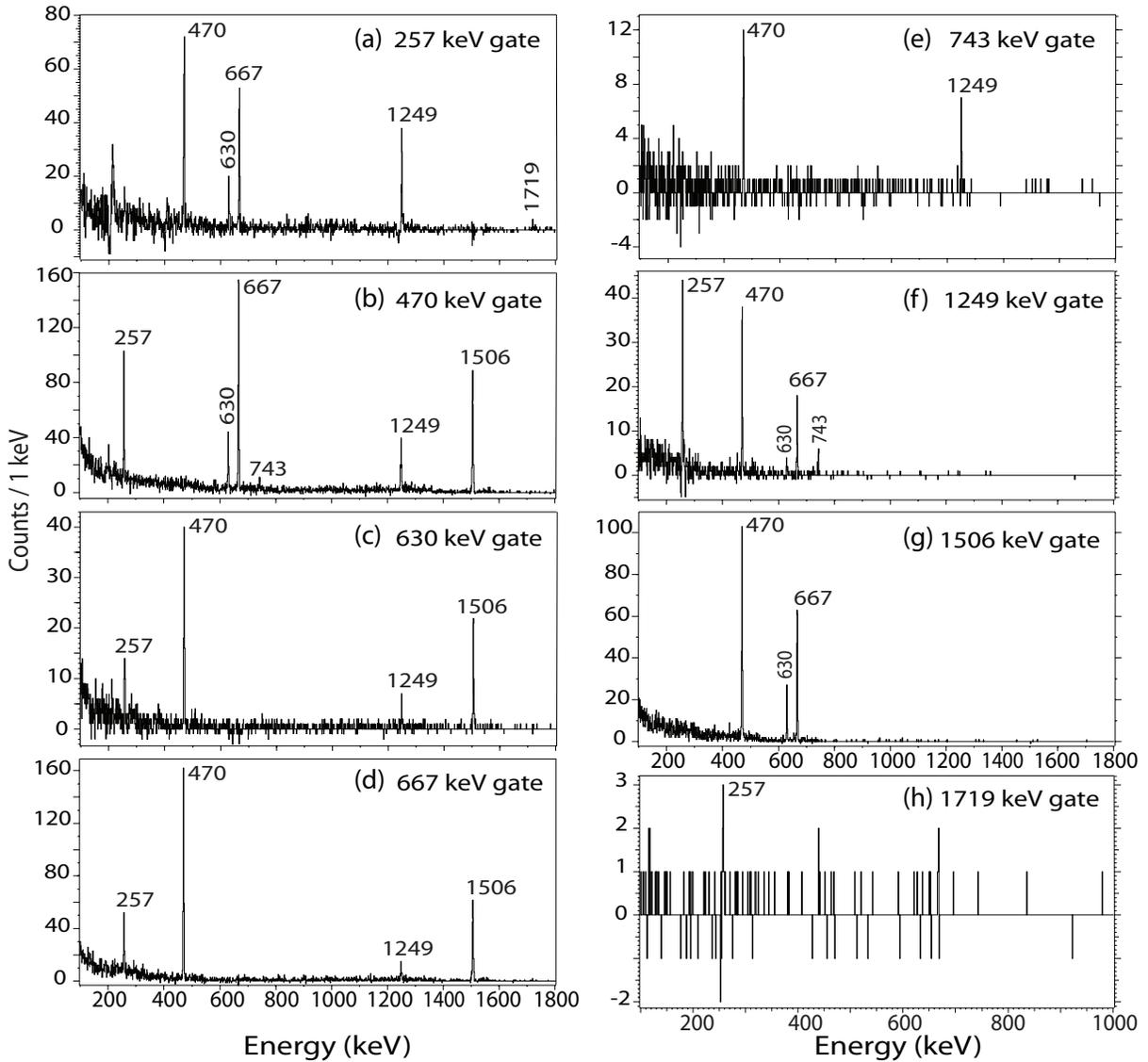}
\caption{Background-corrected fragment-${\gamma}{\gamma}$ coincidence spectra obtained by applying gates on the (a) 257-keV, (b) 470-keV, (c) 630-keV, (d) 667-keV, (e) 743-keV, (f) 1249-keV, (g) 1506-keV and (h) 1719-keV ${\gamma}$ rays. These coincidences were used to deduce the $^{96}$Ag level scheme.}
\label{fig:gg_coinc}
\end{figure*}

\begin{figure}[htpb]
\centering
\includegraphics[width=12cm]{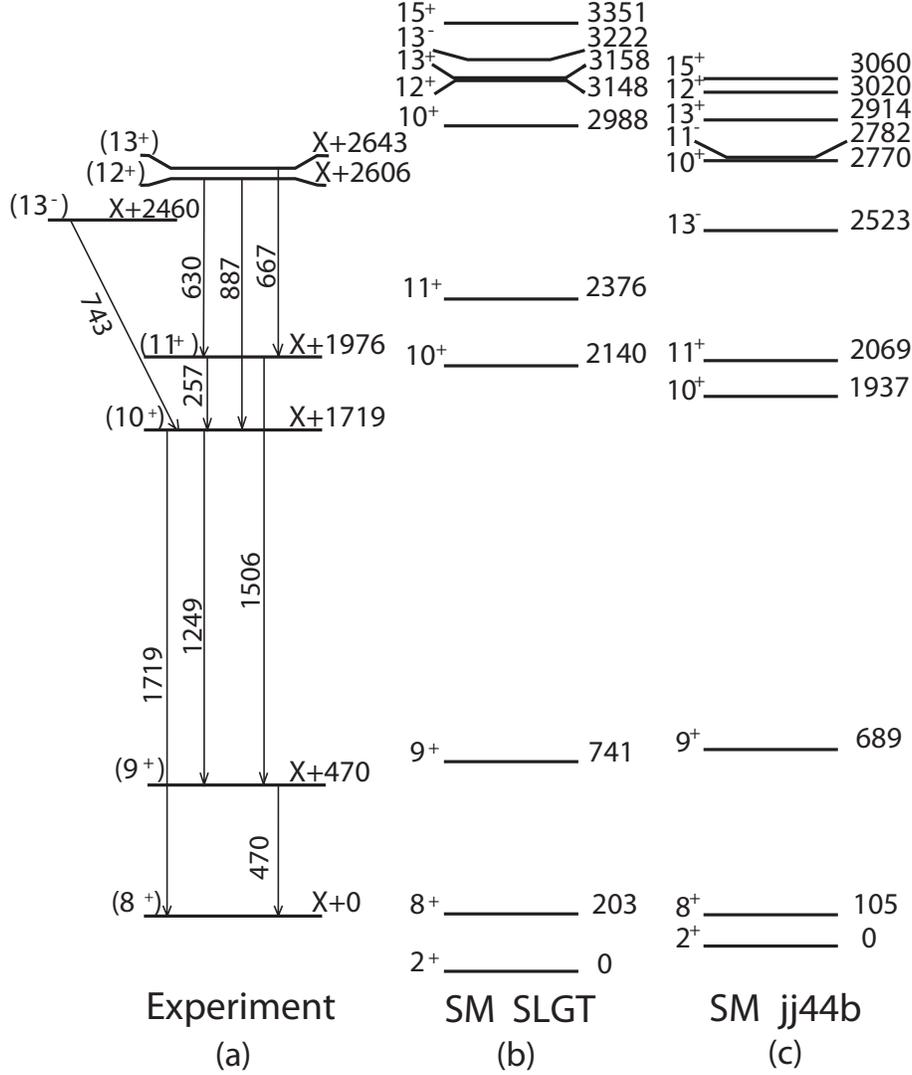}
\caption{(a) Proposed level scheme of $^{96}$Ag, compared to the results of shell-model calculations made with the SLGT interaction (b) and the jj44b interaction (c). Tentative spins and parities were adopted from the results of the jj44b calculation. Deduced experimental level energies (indicated in keV) are relative to the 8$^{+}$ state. The excitation energy of the 8$^{+}$ state relative to the ground state of $^{96}$Ag is not known experimentally. The newly-identified 15$^{+}$ isomeric level is not drawn in the experimental level scheme. See text for details.}
\label{fig:level}
\end{figure}

A 2-dimensional ${\gamma}{\gamma}$ coincidence matrix was constructed for ${\gamma}$ rays which occurred within the first 12 ${\mu}$s after a $^{96}$Ag implantation. 1-dimensional coincidence spectra were then obtained by setting energy gates on the lines of interest in the ${\gamma}{\gamma}$ matrix (Fig. \ref{fig:gg_coinc}). Table \ref{table_gg} summarizes the observed ${\gamma}{\gamma}$ coincidence relationships, as well as the intensities of all transitions, normalized to the transition with an energy of 470 keV. A total of  $1.1 \times 10^{4}$ counts were detected in the 470-keV peak, several orders of magnitude higher than in the previous measurement. Based on these coincidences, their relative intensities, and energy-sum relationships, the decay scheme for $^{96}$Ag presented in Fig. \ref{fig:level} (a) is proposed. The 470-keV line was placed at the bottom of the level scheme based on two observations: it was the most intense transition and it was in coincidence with nearly all other transitions (with exception of the transition with energy 1718 keV). Energy-sum relationships were used to place the 1506-keV transition that is proposed to connect the levels with energies 1976 and 470 keV; and the 1718-keV ground-state transition. The 630-keV and 667-keV ${\gamma}$-rays are not in coincidence with each other, or with the 743-keV line, so they were placed on top of the cascade, both feeding the 1976 keV level.

\begin{table}
\caption{Energies, intensities, and coincidence relationships of ${\gamma}$ rays assigned to the isomeric decay of $^{96}$Ag. \label{table_gg}}
\begin{ruledtabular}
\begin{tabular}{ r r c }
$ E_{\gamma} (keV)$ & $I_{\gamma} (\%)$ & Coincident transitions (keV)\\
\hline
257.2 (2) & 17.3 (16) & 470, 630, 667, 1249, 1719\\
470.0 (2) & 100.0 (67) & 257, 630, 667, 743, 1249, 1506\\
630.1 (2) & 21.8 (20) & 257, 470, 1249, 1506\\ 
667.4 (2) & 71.1 (51) & 257, 470, 1249, 1506\\
742.7 (3) & 4.5 (9) & 470, 1249, 1719\footnote{One count was observed at this energy, which is consistent with the expected number of coinciding ${\gamma \gamma}$ events.}\\
887.4 (6) & 1.0 (4)\footnote{The intensity of this transition was too low to allow observation of ${\gamma \gamma}$ coincidences.}  &    \\
1248.8 (2) & 21.9 (23) &  257, 470, 630, 667, 743\\
1505.9 (2) & 81.4 (63) &  470, 630, 667\\
1718.9 (3) & 1.8 (4) &  257, 743\\
\end{tabular}
\end{ruledtabular}
\end{table}

\begin{figure}[hpb]
\centering
\includegraphics[width=8.5cm]{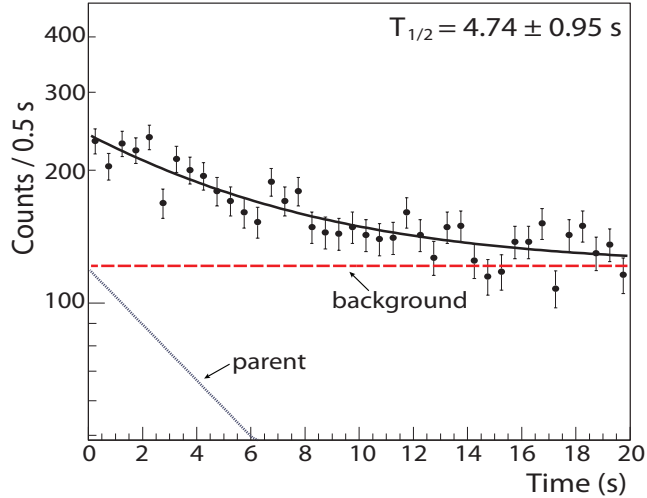}
\caption{(Color Online) ${\beta}$ decay curve of the $^{96}$Ag implants which were correlated with the 470 keV line. The resulting half-life of 4.74(95) s agrees with the known half-life of the ${\beta}$-decaying state in $^{96}$Ag with tentative spin and parity 8$^{+}$.}
\label{fig:betadecay}
\end{figure}

The ${\beta}$ decay of two low-lying states in $^{96}$Ag with proposed spins and parities 2$^{+}$ and 8$^{+}$, and half-lives of 6.9(6) and 4.40(6) s, respectively, was reported by Batist {\it et al.} \cite{Batist2003}. Implanted nuclei correlated with the 470-keV line were tagged to determine which of these two ${\beta}$-decaying states was populated following the decay of the microsecond isomers. The ${\beta}$-decay curve of the tagged nuclei is shown in Fig. \ref{fig:betadecay}. The deduced half-life of 4.74(95) s is consistent with that reported by Batist {\it et al.} for the 8$^{+}$ ${\beta}$-decaying level [4.40(6) s]. Furthermore, the ${\beta}$-delayed ${\gamma}$ spectrum of the tagged nuclei was compared to that of all the $^{96}$Ag implanted nuclei. A considerable decrease of more than 30\% was observed in the relative intensity of the 1415-keV ${\gamma}$-ray transition in the daughter $^{96}$Pd. The 1415-keV transition, according to Batist {\it et al.}, is mostly fed by the ${\beta}$ decay of the 2$^{+}$ level (See Fig. 3 in Ref. \cite{Batist2003}). Therefore, the 1.45 ${\mu}$s isomer most likely populates the 8$^{+}$ ${\beta}$-decaying state. 

The experimental level scheme of $^{96}$Ag is compared with the results of two shell-model calculations in  Fig. \ref{fig:level}. One calculation was performed with the code OXBASH, within the $g_{9/2}-p_{1/2}$ model space using the SLGT effective interaction \cite{Herndl1997} and assuming a $^{100}$Sn core. This calculation predicts one potentially isomeric state with energy of 3148 keV above the 8$^{+}$ state (3351 keV above the 2$^{+}$ state), with spin and parity  15$^{+}$ [see Fig. \ref{fig:level} (b)]. The predicted half-life of the E2 15$^{+}$ ${\rightarrow}$ 13$^{+}$ transition is around 15 ns, much shorter than the half-life of 1.5 ${\mu}$s reported here. The energy of the isomeric transition was estimated as 77~keV using the B(E2) strength from the SLGT calculation and our measured half-life of 1.5~${\mu}$s; such energy is 116 keV smaller than that predicted by the shell model, but well within typical shell model uncertainties. This 77-keV transition is predicted to be highly converted [${\alpha}$=3.67(6)]. No evidence of a 77 keV ${\gamma}$-ray was found in our data, which is consistent with the expected small statistics and signal to background ratio in that energy range. Our experimental setup was also not sensitive to the low-energy electrons expected from this transition.

The second shell-model calculation was carried out in the $(p_{3/2}p_{1/2}f_{5/2}g_{9/2})$ model space considering a $^{56}$Ni core with the jj44b Hamiltonian. The jj44b Hamiltonian was obtained from a fit to about 600 binding energies and excitation energies with a method similar to that used for the JUN45 Hamiltonian \cite{Cheal2010}. Most of the energy data for the fit came from nuclei with $Z=28-30$ and $N=48-50$. With 30 linear combinations of the $J-T$ two-body matrix elements which conserve isospin varied, the rms deviation between experiment and theory for the energies in the fit was about 250 keV. Proton and neutron effective charges of e$_{p}$=1.5 and e$_{n}$=0.5 were used. Predictions obtained with the jj44b Hamiltonian have been compared to experimental data for Ga \cite{Verney2007, Cheal2010} and Cu \cite{Flanagan2010, Vingerhoets2010} isotopes, showing remarkable agreement. The jj44b calculation predicts a 0.99 ${\mu}$s (partial) half-life for a 146 keV 15$^{+}$ ${\rightarrow}$ 13$^{+}$ E2 transition [see Fig. \ref{fig:level} (c)] that feeds a cascade to the 8$^{+}$ level. Again, given the experimental conditions, a 146-keV transition of the expected intensity would not be observed. Additionally, the jj44b calculation predicts a low-lying 13$^{-}$ state, which could explain the longer-lived isomer as a slow E3 transition. Since most of the E3 strength comes from orbitals outside the model space, the reduced E3 transition strength measured in a nearby nucleus was used as reference to estimate the expected lifetime. The closest nucleus to $^{96}$Ag for which a measured E3 value is available is $^{93}$Tc \cite{Brown1978}, with B(E3) = 540 e$^{2}$fm$^{6}$, resulting in a 13$^{-}$ ${\rightarrow}$ 10$^{+}$ half-life of 18 ${\mu}$s. The 13$^{-}$ ${\rightarrow}$ 11$^{+}$ transition is expected to have mixed M2/E3 character, with a partial half-life of a few hundred microseconds, as calculated using the values of the B(M2) and B(E3) obtained in $^{93}$Tc. The small branching ratio for this transition would not have been detected in the present experiment. Therefore, the resulting half-life of the 13$^{-}$ state of 18 ${\mu}$s is within a factor of two of the deduced value of 8.6(63)$~\mu$s reported here.

Overall, both shell model calculations well reproduce the observed level sequence and ${\gamma}$-ray cascades. The jj44b calculation, with its larger model space, predicts excitation energies better than the SLGT one, and reduces the excitation energy of the 13$^{-}$ state sufficiently to provide an explanation for the longer-lived isomer identified in the present work. The jj44b results were used to tentatively assign spins and parities to the experimental levels presented in Fig. \ref{fig:level} (a).

The new data show that the microsecond isomers in $^{96}$Ag are at excitation energies of at least 2.5~MeV and likely have spins and parities of 15$^{+}$ and 13$^{-}$, respectively. With such high excitation energies and spins they are unlikely to be populated in the rp-process.

In summary, a ${\gamma}$-ray cascade was identified from the decay of a 1.45(7)~${\mu}$s isomeric state in the $T_{z}=1$ nucleus $^{96}$Ag. The half-life of the known ${\mu}$s isomer was deduced with higher precision. Evidence for a second ${\mu}$s isomer, along with the ${\gamma}$-ray transitions associated with its decay, was also found. Higher statistics and longer collection times will be needed to determine its half-life to better precision. Both microsecond isomers appear to populate the low-lying, ${\beta}$-decaying 8$^{+}$ state in $^{96}$Ag. A level scheme of the isomeric decay was presented, and was compared with shell model calculations. The shell model calculation employing the jj44b interaction well reproduced the level scheme and isomeric half-lives. Based on our placement of the isomeric states at rather high excitation energies and spins, a non-thermal population in an astrophysical environment that could affect the rp-process is unlikely.

We thank the NSCL operations department for providing the high intensity $^{112}$Sn primary beam and $^{96}$Ag secondary beam. This work is supported by NSF grants PHY08-22648, PHY-06-06007 and PHY-0758099.
\bibliography{Ag96ref}
\end{document}